%% file: paper.tex
\documentclass[aps,prb,reprint]{revtex4-1}
\usepackage{amsmath,amssymb,graphicx,hyperref,xcolor}
\usepackage[utf8]{inputenc}

\begin{document}
\title{Phase transitions and composite order in $\mathrm{U}(1)^N$ lattice London models}

\author{Daniel Weston}
\affiliation{Department of Physics, KTH Royal Institute of Technology, SE-106 91 Stockholm, Sweden}
\author{Karl Sellin}
\affiliation{Department of Physics, KTH Royal Institute of Technology, SE-106 91 Stockholm, Sweden}
\author{Egor Babaev}
\affiliation{Department of Physics, KTH Royal Institute of Technology, SE-106 91 Stockholm, Sweden}

\date{2024-07-31}

\begin{abstract}
The phase diagrams and the nature of the phase transitions in multicomponent gauge theories with an Abelian gauge field are important topics with various physical applications. While an early renormalization-group-based study indicated that the direct transition from a fully ordered to a fully disordered state is continuous for $N = 1$ and $N > 183$, recently it was demonstrated that the transition is discontinuous for $N = 2$. We quantitatively study the dependence on $N$ of the degree of discontinuity of this transition. Our results suggest that the transition is discontinuous at least up to $N = 7$. Furthermore, we demonstrate that, at increased coupling strength, the phase transitions of the neutral and charged sectors of the model split, which for $N > 2$ yields a new phase with composite order. The transition from the composite-order phase to the fully disordered phase is then also discontinuous, at least for $N = 3$ and $N = 4$. Via a duality argument, this indicates that van der Waals-type interaction between directed loops may be responsible for the discontinuous phase transitions in these models.
\end{abstract}

\maketitle
\section{Introduction}

The problem of the order of the superconducting phase transition has long been a subject of intense debate. This debate has been renewed by the consideration of multicomponent superconductors, which currently are of central interest both in superconductivity and as effective field theories for many other condensed matter systems. Especially interesting are new phases with order only in relative degrees of freedom, which can appear in multicomponent superconducting systems.\cite{Svistunov2015, Grinenko2021}

In mean-field BCS theory, the superconducting phase transition is second order,\cite{Ginzburg1950, Bardeen1957b} i.e., there is no latent heat but a discontinuity in the specific heat. The first studies of fluctuations in systems with local $\mathrm{U}(1)$ symmetry included the fluctuation of the gauge field and concluded that -- in contrast to neutral systems, i.e., superfluids -- superconductors have a first-order transition.\cite{Coleman1973, Halperin1974} Later work revised this conclusion by showing that the result may apply only for type 1 superconductors, while to describe the opposite extreme type 2 superconductor it is crucial to include topologically nontrivial excitations, i.e., vortices.\cite{Peskin1978, Dasgupta1981} It was concluded that thermally exited vortex loops make the transition continuous and in the so-called inverted-XY universality class. This conclusion is based on taking the London limit, where a duality mapping was constructed\cite{Peskin1978, Dasgupta1981} that relates a statistical sum of a three-dimensional (3D) lattice London superconductor and a statistical sum of the superfluid 3D XY model with inverted temperature. The duality argument related the statistical mechanics of proliferation of vortices in a system with local $\mathrm{U}(1)$ symmetry, which are directed loops with short-range interaction, to a condensation of directed loops with long-range interaction. Hence the term inverted-XY class: approaching a superconducting transition from below in temperature is dual to approaching a superfluid transition from above. For strongly type-2 superconductors this conclusion was later backed by numerical simulations.\cite{Nguyen1998a} Therefore, for single-component superconductors, a tricritical point determined by the Ginzburg-Landau parameter $\kappa=\lambda/\xi$, which separates discontinuous transitions from continuous ones, was searched for using various methods.\cite{Kleinert1982, Bartholomew1983, Herbut1996, Mo2002, Fejos2016}

The original work of Ref.~\onlinecite{Halperin1974} also considered the case of several complex components. The claim was that the phase transition should be expected to be continuous when the number of components is larger than $N_\mathrm{c} = 183$. The universality class of the transition with multiple components attracted intense interest after the proposal\cite{Senthil2004} that a ``deconfined" quantum phase transition can constitute a direct continuous transition between two states characterized by distinct broken symmetries -- contrary to the reigning ``Landau-Ginzburg-Wilson" paradigm which for such a case insists on a generic discontinuous transition. It has been argued\cite{Senthil2004, Senthil2004b} that phase transitions in quantum antiferromagnets can be mapped onto a phase transition of a $\mathrm{U}(1)\times \mathrm{U}(1)$ superconductor with equal component densities, or of an $\mathrm{SU}(2)$ superconductor. The relationship of the models received numerical backing in Ref.~\onlinecite{Chen2013}. The argument for a direct continuous transition, which was initially rather widely accepted, was based on the idea that since the transition is continuous for $N = 1$ type-2 models \cite{Peskin1978, Dasgupta1981} and models with $N > 183$,\cite{Halperin1974} the transition should be continuous also for $N = 2$. Because the $N = 2$ model is self-dual, a continuous transition would be in a different universality class than inverted XY. However, numerical computations demonstrated that the transition is discontinuous.\cite{Kuklov2005, Kuklov2006, Herland2010} This shifted the search for continuous transitions to $\mathrm{SU}(2)$ models.\cite{Motrunich2008} Subsequently the transition was demonstrated to be discontinuous also in the $\mathrm{SU}(2)$ case, although the discontinuity is weaker.\cite{Kuklov2008b, Kuklov2008, Herland2013} This in turn raised the question of how the phase transitions change with increased numbers of degrees of freedom.

Moreover, even for $\mathrm{U}(1)$-symmetric London systems it has been demonstrated that the phase transition can be discontinuous if the system has several components. Namely, in Ref.~\onlinecite{Sellin2016}, it was found that if one takes a model with $\mathrm{U}(1)^N$ symmetry and breaks the symmetry explicitly to $\mathrm{U}(1)$ by adding Josephson terms, there is a tricritical point and there appears a discontinuous transition (see also Ref.~\onlinecite{Galteland2016c}). Since there is only one phase transition in a $\mathrm{U}(1)$ system, and only one type of directed loop with short-range interaction that can proliferate, the existence of a tricritical point shows that the form of this interaction is important, in contrast to the usual assumption in duality mappings. In the directed-loops model, directly modifying the short-range interaction potential by adding short-range attractive parts has also been shown to lead to a discontinuous transition under certain conditions.\cite{Meier2015}

The origin of the discontinuous phase transitions in multicomponent systems is still poorly understood. It remains an outstanding open question whether there exist multicomponent gauge theories with Abelian gauge fields that have continuous phase transitions or if such theories can be generically ruled out or ruled out for a class of models. This motivated recent works that have embarked on renormalization group studies of theories with higher $N$.\cite{Fejos2016, Fejos2017, Nogueira2013, Ihrig2019} Exploration of related questions in a more formal framework is carried out in Ref.~\onlinecite{Gorbenko2018}.

Importantly, at strong intercomponent coupling gauge theories have been shown to have multiple phase transitions and to form new phases with so-called composite order, which in the context of two-component bosonic lattice systems are also called paired phases.\cite{Babaev2002, Babaev2004b, Smorgrav2005b, Kuklov2006, Kuklov2008b, Bojesen2014a, Svistunov2015} These new phases occur when the phase transition splits into two: at lower temperature the phase sum of all the components disorders, the Meissner effect disappears, and the system enters a phase with order only in phase differences between components. Recently, an observation of a related phase was reported above the superconducting transition in a $\mathrm{U}(1)\times Z_2$ superconductor.\cite{Grinenko2021}

There is a mean-field argument\cite{Kuklov2006} that relates the aforementioned discontinuous phase transitions to the presence of composite-order phases. The mean-field argument indicates that the direct transition from the fully ordered phase to the fully disordered phase is discontinuous for some range of couplings. However, numerical calculations in the $\mathrm{U}(1)\times \mathrm{U}(1)$ and $\mathrm{SU}(2)$ models that employ the flowgram method show that the direct transition is always discontinuous, not only in proximity to the composite-order phase.\cite{Kuklov2006, Kuklov2008b, Kuklov2008, Herland2013} Also, multicomponent superconducting models without composite-order phases can exhibit discontinuous transitions.\cite{Sellin2016}

In Ref.~\onlinecite{Sellin2016} it was conjectured that van der Waals-type interaction between directed composite loops is responsible for driving the phase transitions to be discontinuous. This was backed by relating the tricritical point of Josephson-coupled systems to the range of the van der Waals-like forces between composite vortices, as the composite vortices can be viewed as bound states of electrically charged strings. The conjecture was also backed by the numerical observation that, at least for low $N$, the degree of discontinuity in a Josephson-coupled system may, under certain conditions, increase with $N$.

The scenario where van der Waals-type interaction between directed composite loops drives the discontinuous phase transitions deviates from the mean-field analysis in the following testable aspect: The mean-field analysis\cite{Kuklov2006} predicts that the transition from the fully ordered phase to the composite-order phase should be discontinuous, at least near the bicritical point, while the transition from the composite-order to the disordered phase should be continuous. However, by a duality mapping the transition from the disordered to the composite-order phase with decreased temperature can be mapped onto proliferation of directed composite loops that should also interact via van der Waals-type forces that can drive the transition discontinuous.

The goal of this paper is to numerically investigate phase transitions in $\mathrm{U}(1)^N$ London models of $N$ complex fields coupled to a noncompact Abelian gauge field for $N > 2$. The paper is organized as follows. First we present the models that we consider. Then we describe the numerical methods that we use, including the observables used to locate and characterize phase transitions. Finally, we present and discuss our results on phase diagrams and the nature of certain transitions.

\section{Models}

We consider $\mathrm{U}(1)^N$-symmetric London models with identical components in three spatial dimensions, given by the Hamiltonian density
\begin{equation}
  h = \tfrac{1}{2}(\nabla \times \mathbf{A})^2
      + \tfrac{1}{2} \sum_i |(\nabla + \mathrm{i}q\mathbf{A}) \psi_i|^2.
  \label{contf}
\end{equation}
Here $\mathbf{A}$ is the magnetic vector potential and the $\psi_i$ are matter fields corresponding to the superconducting components. For each $N$ the amplitudes $|\psi_i| = 1/\sqrt{N}$, so that the total superconducting density $\sum_i |\psi_i|^2 = 1$.

The model (\ref{contf}) can be rewritten in terms of neutral and charged modes as \cite{Babaev2002b}
\begin{equation}
  h = \tfrac{1}{2} \mathbf{j}^2 +
      \tfrac{1}{2} \sum_{i<j} |\psi_i|^2 |\psi_j|^2 (\nabla \phi_{ij})^2 +
      \tfrac{1}{2} (\nabla \times \mathbf{A})^2,
  \label{modes}
\end{equation}
where $\phi_{ij} = \phi_j - \phi_i$ and $\mathbf{j}$ is the density of charged supercurrent:
\begin{equation}
  \mathbf{j} = \sum_i |\psi_i|^2 (\nabla \phi_i + q \mathbf{A}).
  \label{current}
\end{equation}
Note that the nonmagnetic energy can be divided into a term that gives the energy from electrically charged currents and a set of terms that give the energy from electrically neutral currents consisting of counterflows of charged condensates. Hence the model has one charged mode and $N-1$ neutral modes [or, more properly, an $(N-1)$-dimensional space of neutral modes]. Importantly, the original fields $ \phi_i$ are $2\pi$ periodic.

Because of the coupling to vector potential, vortices that have winding in the phase of each component (composite vortices) will have finite energy per unit length,\cite{Babaev2002b} as do vortices in ordinary single-component superconductors. On the other hand, vortices that do not have phase winding in each component, e.g., those that have winding in only one component (fractional vortices), will have an energy per unit length that diverges logarithmically with system size, as in ordinary single-component superfluids. Composite vortices carry magnetic flux equal to the ordinary superconducting flux quantum, whereas fractional vortices carry only a fraction of a flux quantum. This fraction is equal to the ratio of the density $|\psi_j|^2$ of the component $j$ in question to the total density $\sum_i |\psi_i|^2$. See detailed discussion of the vortex solutions in Ref.~\onlinecite{Babaev2002b} and Chap.~6 of Ref.~\onlinecite{Svistunov2015}.

We denote a type of vortex in an $N$-component superconductor by a tuple of $N$ integers, where the $i$th integer gives the winding of the phase of the $i$th component. For example, in the four-component case, a fractional vortex with phase winding in only the first component is denoted $(1,0,0,0)$, while a composite vortex with winding in each component is denoted $(1,1,1,1)$.

\section{Monte Carlo simulation methods and observables}

We discretize the model (\ref{contf}) on a three-dimensional simple cubic lattice with $L^3$ sites and lattice constant $a = 1$. The discretized model is given by the Hamiltonian density
\begin{equation}
  h = \tfrac{1}{2} \sum_{k<l} F_{kl}^2
      - \sum_{i,k} |\psi_i|^2 \cos \chi_{i,k}(\mathbf{r}),
  \label{lattf}
\end{equation}
where
\begin{equation}
  F_{kl} = A_k(\mathbf{r}) + A_l(\mathbf{r} + \mathbf{k})
           - A_k(\mathbf{r} + \mathbf{l}) - A_l(\mathbf{r})
\end{equation}
is a lattice curl,
\begin{equation}
  \chi_{i,k}(\mathbf{r}) =
    \phi_i(\mathbf{r}+\mathbf{k}) - \phi_i(\mathbf{r}) + q A_k(\mathbf{r})
\end{equation}
is a gauge-invariant phase difference, $k$ and $l$ signify coordinate directions, and $\mathbf{k}$ is a vector pointing from a lattice site to the next site in the $k$ direction. We use periodic boundary conditions in all three spatial directions. The thermal probability distribution for configurations of the system at inverse temperature $\beta$ is given by the Boltzmann weight
\begin{equation}
  \mathrm{e}^{-\beta H}, \quad H = \sum_{\mathbf{r}} h(\mathbf{r}),
\end{equation}
and we generate representative samples from these thermal distributions using Monte Carlo simulation.

The simulations are performed using the Metropolis-Hastings algorithm with local updates of each of the degrees of freedom. At each point in the lattice the three components of the vector potential are updated either together or one at a time and the phases are updated one at a time. We also use parallel-tempering swaps between systems with neighboring temperatures. We typically use $32$ or $64$ parallel temperatures. The temperatures are adjusted within a fixed interval during the equilibration in order to make the acceptance ratios for parallel-tempering swaps equal for all neighboring pairs of temperatures. Equilibration is checked by comparing results obtained using the first and second halves of the data gathered after equilibration, by comparing inverse-temperature derivatives obtained using finite differences and statistical estimators, and by inspecting obtained time series of Monte Carlo data.

We use Ferrenberg-Swendsen reweighting\cite{Ferrenberg1988} in order to precisely determine peak values of heat capacity. We do this by reweighting from the simulated temperature that gives the largest value of the heat capacity.

Errors are estimated by bootstrapping and stated errors correspond to one standard error. The bootstrapping procedure we use is as follows. First, note that when a parallel-tempering swap occurs, the state corresponding to either one of the involved temperatures is immediately changed to a completely new state that is uncorrelated with the previous state. Nonetheless, the state for a given temperature immediately prior to the swap may, due to further swaps, be correlated with a subsequent state for the same temperature. The sequence of states for a given temperature will therefore be such that there is correlation between states at different times that is not mediated by the states at intermediate times. Consequently, prior to blocking the data series should be reordered so that values from correlated states are grouped together (with the order within such a group being preserved). Having performed this reordering, we divide the data into at least 20 blocks, each of which is at least 20 times as long as the longest of the autocorrelation times (estimated using the reordered series) for the energy and the quantities we measure in order to locate phase transitions. From these blocks we generate at least 100 resampled data sets, each of which gives a value for the observable in question. Finally, we estimate the standard error $\sigma_x$ in the observable $x$ as
\begin{equation}
  \sigma_x = \sqrt{\frac{1}{n-1} \sum_{i=1}^n (x_i - \bar{x})^2},\quad
  \bar{x} = \frac{1}{n} \sum_{i=1}^n x_i,
\end{equation}
where $n$ is the number of resamplings and $x_i$ is the value of $x$ obtained from resampling number $i$.

We now describe quantities that are measured during the simulations and the methods we use to locate and characterize phase transitions.

\subsection{Locating superconducting transitions}

For the purpose of locating superconducting transitions we consider the dual stiffness\cite{Motrunich2008, Herland2013, Carlstrom2015}
\begin{equation}
  \rho^\mu(\mathbf{q}) =
    \left\langle \frac{\left| \sum_{\mathbf{r},\nu,\lambda}
      \epsilon_{\mu\nu\lambda} \Delta_\nu A_\lambda(\mathbf{r})
      \mathrm{e}^{\mathrm{i}\mathbf{q} \cdot \mathbf{r}} \right|^2}
      {(2\pi)^2 L^3} \right\rangle,
\end{equation}
where $\epsilon_{\mu\nu\lambda}$ is the Levi-Civita symbol, $\Delta_\nu$ is a difference operator, and $\langle \cdot \rangle$ is a thermal expectation value. More precisely, we consider the dual stiffness in the $z$ direction evaluated at the smallest relevant wave vector in the $x$ direction $\mathbf{q}_\mathrm{min}^x = (2\pi/L, 0, 0)$, i.e., $\rho^z(\mathbf{q}_\mathrm{min}^x)$. This quantity, which we denote simply as $\rho$, will in the thermodynamic limit approach zero in the superconducting phase in which the Meissner effect suppresses fluctuations of the magnetic field. In nonsuperconducting phases $\rho$ will have a finite value. Consequently, it is a dual order parameter in the sense that it is zero in the ordered phase and nonzero in the disordered phase.

Generally, our estimates of superconducting critical temperatures are given by finite-size crossings of the quantity $L\rho$ (dual stiffness scaled by system size), extrapolated to the thermodynamic limit.\cite{Motrunich2008, Herland2013} (Although the transitions we study are not necessarily continuous, we expect that such crossings will converge to the critical temperature.) We estimate the locations of the crossings by simple linear interpolation between simulated inverse temperatures. For some of the transitions we study, the crossings are not sufficiently well behaved for extrapolation to the thermodynamic limit to be possible. Therefore, we also consider the inflection points of $\rho(\beta)$, again extrapolated to the thermodynamic limit. We estimate the location of such an inflection point by reweighting from the simulated inverse temperature with the largest absolute value of the inverse-temperature derivative. In some cases, this second method does not work, either because the quality of the data is insufficient or because extrapolation to the thermodynamic limit is not possible. In those cases in which both methods appear to work, the results are consistent.

\subsection{Locating superfluid transitions}

In order to locate superfluid transitions we use the helicity modulus, which measures the global phase coherence of a superfluid. More precisely, it measures the cost in free energy of imposing an infinitesimal phase twist. Imposing a twist in a certain linear combination $\sum_i a_i \phi_i$ of the phases amounts to replacing the phase $\phi_i$ by
\begin{equation}
  \phi_i'(\mathbf{r}) = \phi_i(\mathbf{r}) -
                        a_i\,\boldsymbol{\delta}\cdot\mathbf{r}.
\end{equation}
For a given linear combination of phases, the helicity modulus is by definition the second derivative
\begin{equation}
  \Upsilon_{\mu,\{a_i\}} = \frac{1}{L^3} \frac{\partial^2 F[\phi_i']}{\partial \delta_\mu^2},
\end{equation}
where $F$ is the free energy. By considering the fundamental relations
\begin{equation}
  F = -T\ln Z, \quad Z = \mathrm{Tr}\,\mathrm{e}^{-\beta H},
\end{equation}
one can derive an expression for the helicity modulus $\Upsilon_{\mu,\{a_i\}}$ in terms of the first and second derivatives of the Hamiltonian $H$ with respect to the phase-twist parameter $\delta_\mu$. We use this expression to evaluate helicity moduli from our numerical simulations. How this is done is described in more detail in Ref.~\onlinecite{Weston2021}. (However, here we do not always use reweighting to improve our helicity-modulus data.)

Our estimates of superfluid critical temperatures are given by finite-size crossings of the quantity $L\Upsilon$ (helicity modulus scaled by system size), extrapolated to the thermodynamic limit. This is similar to how we use finite-size crossings of $L\rho$ in the superconducting case. Again, we use linear interpolation between simulated inverse temperatures to estimate the locations of the crossings.

\subsection{Characterizing discontinuous transitions}

For a system that has a temperature-driven discontinuous phase transition, the distribution of internal energies will be bimodal for large enough system sizes. This is a consequence of phase coexistence, which is characteristic of discontinuous transitions. The occurrence of energy distributions with bimodality that becomes increasingly pronounced with increasing system size is strong evidence that a transition is discontinuous.

Apart from assessing whether a transition \emph{is} discontinuous, we estimate the \emph{degree} of discontinuity by considering the finite-size scaling of the heat capacity. For a discontinuous transition in a $d$-dimensional system of linear size $L$ the heat-capacity maximum is expected to scale as $c_\mathrm{max}\sim L^d$ in the thermodynamic limit.\cite{Challa1986} Measuring the heat-capacity maximum for large $L$ and fitting the curve $c_\mathrm{max} = k L^d + m$ gives a measure $k$ of the strength of the discontinuous transition. The quantity $k$ is a measure of the (square of the) latent heat that is normalized by the transition temperature.

\section{Results and discussion}

\subsection{Phase diagrams}

\begin{figure}
  \includegraphics{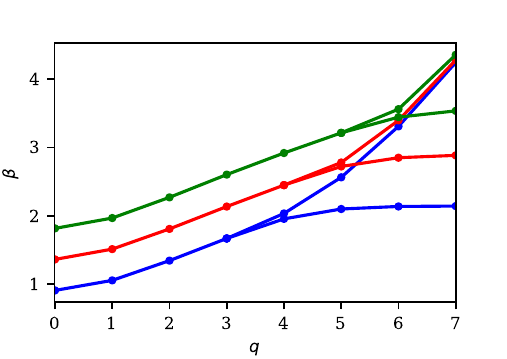}
  \caption{Phase diagrams for $N = 2$ (blue, lower diagram), $N = 3$ (red, middle diagram), and $N = 4$ (green, upper diagram). In each case there is a direct transition from a fully ordered to a fully disordered state for small enough electric charge $q$, whereas there are two separate transitions for large enough $q$. The phase between the two separate transitions is a state with composite order, in which there is order only in phase differences. This is a superfluid state where only counterflow of components is dissipationless. With increased $N$ the area of the composite-order phase shrinks, since the bicritical point that separates these two regimes moves to higher electric charge. Errors are estimated to be smaller than symbol sizes and lines are a guide to the eye.}
  \label{phase_diag}
\end{figure}

We begin by calculating the phase diagrams for $N = 2$, $N = 3$, and $N = 4$, which are shown in Fig.~\ref{phase_diag}. Consider, as an example, the case $N = 4$. As the coupling constant increases, the energy of a composite vortex $(1,1,1,1)$ decreases relative to the energy of fractional vortices such as $(1,0,0,0)$. At a certain coupling strength, proliferation of composite vortices takes place at a significantly lower temperature than the one required to proliferate fractional vortices. In this case, the system enters a composite-order phase characterized by order only in phase differences between components. This phase is described by the effective Hamiltonian
\begin{equation}
  h = \tfrac{1}{2} \sum_{i<j} \varrho (\nabla \phi_{ij})^2.
  \label{modes2}
\end{equation}
While for $N = 2$ this state has the physical interpretation that only counterflow is dissipationless, in general it describes $N-1$ neutral modes involving $N$ phases. The $N > 2$ case represents a new kind of superfluid state in which the dissipationless counterflow allows exchange of particles between the counterflowing components. This illustrates that while the order parameter characterizing this phase is a product of the original fields $\psi_i$, this state cannot be interpreted as a real-space pairing. Such states are currently of significant interest following the experimental report of a discrete-symmetry example of such composite order.\cite{Grinenko2021}

Note that with increased $N$ the relative volume of the composite-order phase on the phase diagram decreases. This is because with increased $N$ at fixed total density $\sum_i |\psi_i|^2 = 1$, the energy of fractional vortices such as $(1,0,0,0)$ becomes smaller due to the diminishing fraction of the flux quantum carried by an elementary vortex and the diminishing prefactor for the logarithmically divergent part of the energy, while the energy of a composite vortex $(1,1,1,1)$ does not change. (Note that the temperature of the superconducting transition saturates in the limit of increased charge to a value that only weakly depends on $N$.) Therefore, with increased $N$ a higher coupling constant is required to substantially split the temperatures of the superfluid and superconducting transitions.

Finally, note that at charge $q = 0$ the model consists of $N$ uncoupled XY models. Thus the critical inverse temperature at $q = 0$ is trivially proportional to $N$ as a consequence of the total superconducting density being held fixed, as this implies that the prefactors of individual cosine terms scale as $1/N$.

\subsection{Degree of discontinuity}

Having established the phase diagram, we move to examining the nature of the phase transitions. We focus first on direct transitions from the fully ordered phase to the fully disordered phase and choose to consider the fixed coupling constant $q = 2$. For each $N$ in the range from 2 to 7, these transitions show signs of being discontinuous. For $N = 2,3,4$ we observe clear evidence of discontinuity in the form of bimodality in the energy distributions that becomes more pronounced with increasing system size. For $N = 5,6$ we observe bimodality for the largest systems we simulate. For $N = 7$ we observe apparent precursors of bimodality: anomalously flat-peaked energy distributions in the vicinity of the transition temperature, with distributions skewed toward higher and lower energies for lower and higher temperatures, respectively.

Assuming that the transitions in question are discontinuous, we consider their degree of discontinuity as measured by the quantity $k$. Note that considering a fixed coupling constant will tend to lead to underestimation of how pronounced the discontinuous phase transitions are for larger values of $N$. This is because the transitions are typically most discontinuous close to the bicritical point. With increased $N$ the composite-order phase shrinks and thus one effectively moves away from the bicritical point.

\begin{figure*}
  \includegraphics{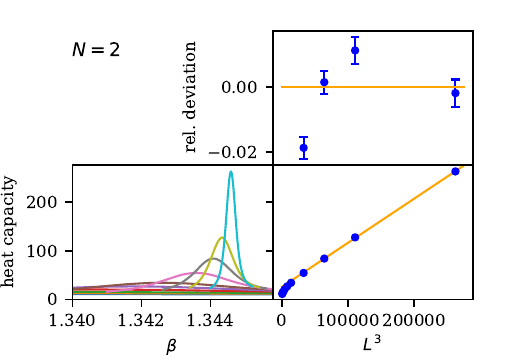}
  \includegraphics{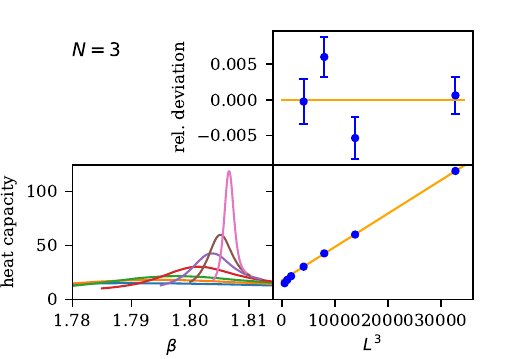}
  \includegraphics{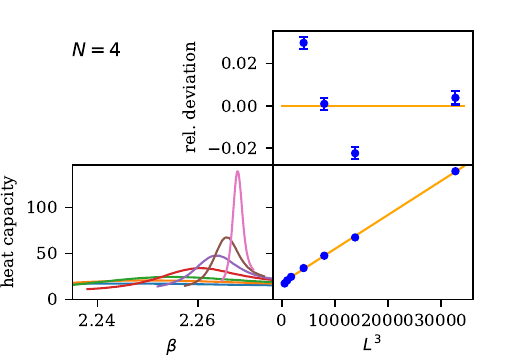}
  \includegraphics{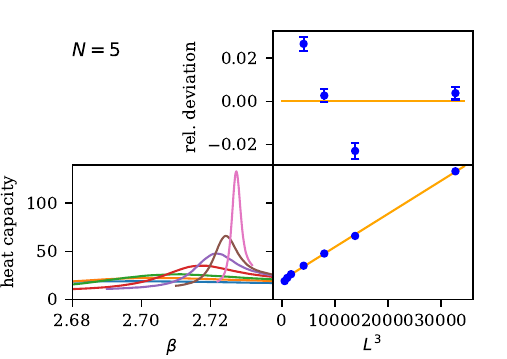}
  \includegraphics{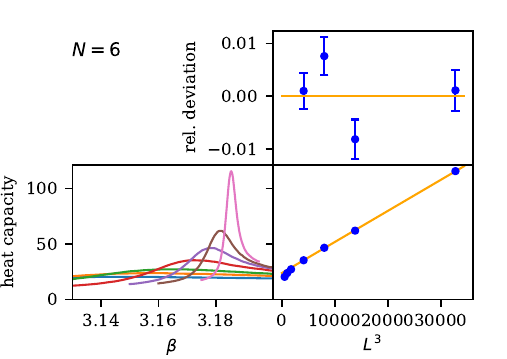}
  \includegraphics{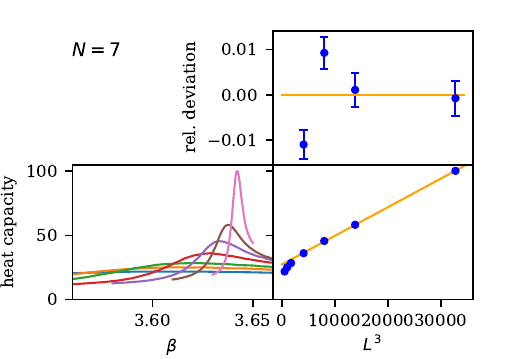}
  \caption{The strength of a discontinuous transition is quantified here by the asymptotic slope $k$ of the peak value $c_\mathrm{max}$ of the heat capacity $c = L^{-3}\,\mathrm{d}\langle E \rangle/\mathrm{d}T$ versus the system volume $L^3$. Finite-size scaling analyses for the cases $N=2$--$7$ with $q = 2$ are shown. We estimate the slope $k$ by fitting a line $c_\mathrm{max} = kL^3 + m$ for the larger systems. The system sizes are $L = 8, 10, 12, 16, 20, 24, 32, 40, 48, 64$ for $N = 2$ and the same sizes in the range $L = 8$--$32$ for larger $N$. In each case we estimate $k$ using the four largest system sizes. For each fitted line, the relative deviation of the corresponding data points from the line is shown, together with the estimated relative errors in these peak heat-capacity values.}
  \label{first}
\end{figure*}

\begin{figure}
  \includegraphics{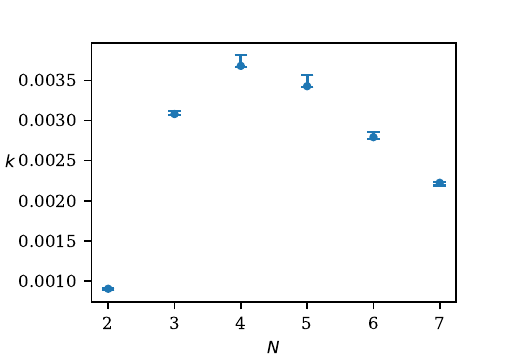}
  \caption{Estimates of the strength $k$ of the discontinuous phase transitions at fixed coupling constant $q = 2$ as a function of the number $N$ of components. There is an increase of $k$ up to $N = 4$ and then a decrease with $N$. The error estimates (which are asymmetric) take into account both the statistical uncertainty and the difference in results when using only the two or three largest system sizes instead of the four largest.}
  \label{trend}
\end{figure}

The process of estimating $k$ is illustrated in Fig.~\ref{first} and the resulting values are shown in Fig.~\ref{trend}. Note that the data points that the lines are fitted to do not all fit the lines within the statistical uncertainty. Therefore, in Fig.~\ref{trend}, we have chosen to display error bars that not only reflect the statistical uncertainty, but also the difference in results obtained when using only the two or three largest system sizes instead of the four largest.

Before we discuss these results, the following additional caveat should be made. Whether system sizes used in a simulation of a discontinuous transition are in the finite-size-scaling regime can be assessed by the criterion\cite{Lee1991} that the free-energy barrier both (1) be much larger than unity and (2) scale as $L^{d-1}$. The system sizes we use do not fulfill this criterion; in fact, they do not even fulfill the first part of the criterion. Nevertheless, we believe that the data presented here may give a meaningful indication of the degrees of discontinuity.

Since the transition in question is argued to be continuous for large enough $N$, it was widely expected that there would be a monotonic decrease of the degree of discontinuity with $N$. However, the results in Fig.~\ref{trend} suggest that this is not the case. Instead, at least for small $N$ the degree of discontinuity $k$ appears to increase with $N$, despite the fact that at fixed $q$ we are getting further away from the bicritical point with increased $N$.

Since for the case of uncoupled components ($q = 0$) the heat capacity is trivially proportional to $N$ (regardless of the total density), one may reasonably ask whether $k/N$ is a more appropriate measure of the degree of discontinuity than $k$ itself. However, note that even if one considers $k/N$ the transition for $N = 3$ is more discontinuous than that for $N = 2$. As mentioned, this is true despite the fact that we effectively move away from the bicritical point. This strengthens the conclusion that the transitions become more discontinuous with increasing $N$, at least for small values of $N$.

As the quantity $k$ is proportional to the \emph{square} of the latent heat, one may also reasonably ask whether $\sqrt{k}$ is a more appropriate measure of the degree of discontinuity than $k$ itself. In total, we thus have four proposed measures of the degree of discontinuity: $k$, $k/N$, $\sqrt{k}$, and $\sqrt{k}/N$. For each of these measures, our results suggest that the transition is significantly more discontinuous for $N = 3$ than for $N = 2$.

\subsection{Discontinuous superfluid transitions}

Finally, we test another aspect of the hypothesis that van der Waals-type forces between directed loops are a ``microscopic" reason why phase transitions in multicomponent gauge theories are discontinuous. In terms of vortex proliferation, both the direct transition and the superconducting transition from the fully ordered to the composite-order phase involve proliferation of composite integer-flux vortex loops. Because composite vortices can be viewed as bound states of electrically charged strings, they have van der Waals attractive forces, potentially leading to phase separation of vortex tangles. The superfluid transition from the composite-order to the disordered phase is driven by noncomposite fractional vortex loops such as $(1,0,0,...)$ in a background of proliferated composite vortices $(1,1,1,...)$. However, in the dual picture the same transition can be mapped to proliferation of directed composite loops if one approaches the transition from the disordered phase.\cite{Svistunov2015} The hypothesis thus also leads to the expectation that the superfluid transition is discontinuous.

\begin{figure}
  \includegraphics{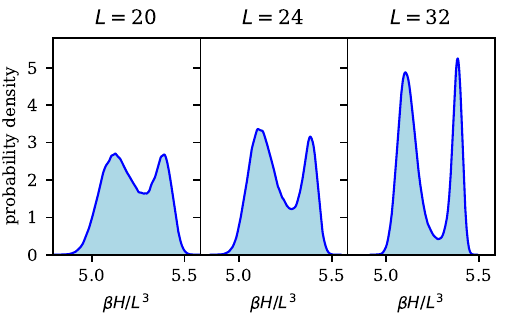}
  \caption{Distributions of action density $\beta H/L^3$ for temperatures in the vicinity of the superfluid transition for $N = 3$ and $q = 6$; the temperatures are chosen so that the peaks are as near as possible equal in height. The bimodality becomes more pronounced with increasing system size, which shows that the transition is discontinuous.}
  \label{hist}
\end{figure}

We demonstrate that the superfluid transition \emph{is} discontinuous, at least for $N = 3$ and $N = 4$, and at least close to the bicritical point. Our evidence for this is the bimodality of energy distributions, which becomes more pronounced with increasing system size. As an example we show distributions of action density $\beta H/L^3$ for the superfluid transition for $N = 3$ and $q = 6$ for the system sizes $L = 20, 24, 32$ (Fig.~\ref{hist}). All the superfluid transitions in the phase diagrams for $N = 3$ and $N = 4$ in Fig.~\ref{phase_diag} show such bimodality of energy distributions.

\section{Conclusion}
The nature of the phase transitions in multicomponent theories coupled to a noncompact Abelian gauge field is an outstanding question. While the $\mathrm{U}(1)\times \mathrm{U}(1)$ case is well investigated numerically, in this paper we have addressed the question of how the phase diagrams and nature of the phase transitions evolve with increasing number of components in $\mathrm{U}(1)^N$-symmetric London models.

For all numbers of components for which we have determined phase diagrams ($N = 2,3,4$), we have established the presence of a phase with composite order, in which order exists only in phase differences. At the same time, the size of the composite-order phase shrinks with increasing $N$.

Based on renormalization group calculations it has been claimed that for sufficiently large $N$ the direct transition may become continuous.\cite{Halperin1974, Ihrig2019} Nonetheless, we have found indications that the transition may become \emph{more} discontinuous with increasing $N$, at least for small $N$. We have also seen indications that the transition is discontinuous at least up to $N = 7$.

Finally, we have demonstrated that, in contrast to previous expectations, the superfluid transition (i.e., the transition from the composite-order to the fully disordered phase) is also discontinuous, at least for certain values of the coupling constant for $N = 3$ and $N = 4$. This suggests that van der Waals-like interaction between directed composite loops may be an important factor for the discontinuous character of phase transitions in this kind of multicomponent gauge theory.

\begin{acknowledgments}
The work was supported by the Swedish Research Council Grants No.\ 642-2013-7837, No.\ 2016-06122, No.\ 2018-03659, and No.\ 2022-04763, and by the G\"oran Gustafsson Foundation for Research in Natural Sciences and Medicine. The computations were enabled by resources provided by the National Academic Infrastructure for Supercomputing in Sweden (NAISS), partially funded by the Swedish Research Council through Grant Agreement No.\ 2022-06725.
\end{acknowledgments}

\input{paper.bbl}

\end{document}

%% file: paper.bbl
%